# Analysis and Experimental Validation of the WPT Efficiency of the Both-Sides Retrodirective System

**Charleston Dale Ambatali** [1,2,*], **Shinichi Nakasuka** [1], **Bo Yang** [3], **and Naoki Shinohara** [3]

[1] Department of Aeronautics and Astronautics, The University of Tokyo, Japan

[2] Electrical and Electronics Engineering Institute, University of the Philippines, Philippines

[3] Research Institute for Sustainable Humanosphere, Kyoto University, Japan

\* Correspondence: cmambatali@up.edu.ph

**Abstract:** The retrodirective antenna array is considered as a mechanism to enable target tracking of a power receiver for long range wireless power transfer (WPT) due to its simplicity in implementation using only analog circuits. By installing the retrodirective capability on both the generator and rectenna arrays, a feedback loop that produces a high efficiency WPT channel is created. In this paper, we characterize the dynamics of this phenomenon using a discrete-time state-space model based on S-parameters and show that the system can naturally achieve maximum theoretical WPT efficiency. We further confirmed the theoretical analysis through a hardware experiment using a 12-port circuit board with measurable S-parameters mimicking a static wireless channel. The results collected from the hardware experiment show agreement with the proposed theoretical framework by comparing the theoretical efficiency with the measured efficiency and by showing that the collected data points follow the predicted condition to achieve maximum efficiency.

## 1. Introduction

Long range wireless power transfer (WPT) envisions a system with multiple antennas on both the receiver and the transmitter to make the whole antenna system large enough for beam control or beam focusing. An estimate of the total antenna array size can be made by using the maximum distance of the radiative near-field zone [1]. Let $\lambda$ be the wavelength of operation which is related to the wave frequency, $f$, and the speed of light in free space, $c$, by $c = \lambda f$. If $D$ is the antenna diameter, then the maximum distance can be calculated by $2D^2/\lambda$ [2]. Alternatively, a beam collection efficiency (BCE) approach [3, 4] can also be used for estimating antenna size. This is estimated by the radiation pattern of the antenna array design given the total area of the power receiver.

According to the limit of the radiative near-field, in the scale of space-based solar power (SBSP) from geostationary orbit (~36,000 km distance), a minimum antenna length of 1 km is needed to be in the radiative near-field [5]. Using the BCE approach, we have also made an estimate that a 2.5 km×2.5 km antenna array in orbit can focus a beam to a 5 km diameter on the ground for a high beam efficiency [6]. For a spacing of half a wavelength between antenna elements, this corresponds to a minimum of 1.5 billion to approximately 10 billion antennas.

To generate a highly efficient WPT beam, precise phase and amplitude control must be performed on each antenna element. Multiple techniques have been developed for this problem. One approach uses the position tracking sensors (e.g. image recognition, GPS) to calculate the optimal antenna setting [7]. Another iteratively calculates for the optimal setting by collecting data the power output at the receiver array for every setting. This is possible by using telemetry [8, 9]. Problems arising from these approaches are its scalability towards larger systems like SBSP mentioned previously.

In general, these methods show a trade-off between implementation complexity, efficiency performance, and convergence time. A simple implementation with fast convergence time suffers from sub-optimal efficiency. On the other hand, trying to achieve optimal efficiency requires complex implementation and slower convergence time. To avoid this inherent trade-off problem, retrodirective antennas [10] are considered since they can be implemented by purely analog circuits [11]. This system can track the source of a signal and send the power towards the same direction almost instantly without the need for complex calculations. By equipping retrodirective capability both the generator array and the receiver array, a feedback loop is created that creates an optimal beam in the space between the antenna arrays as long as marginal stability is maintained [12]. This is called the both-sides retrodirective antenna array (BS-RDAA) system. In this paper, we present a complete analysis of the behavior of this system that has been lacking from the previously published study [12]. We show that the maximum WPT efficiency (WPTE) can be achieved at a generalized marginal stability condition determined by the inherent properties of the wireless channel from its S-parameter matrix. Moreover, we present an experiment setup that aims to demonstrate the accuracy of the theoretical analysis performed. The results show good agreement between theory and experiment.

In Section 2, we discuss in more detail the different works that have been done on power beam control. We also briefly discuss the BS-RDAA concept positioned among them. From there, we present a framework towards analyzing WPTE in Section 3 from the perspective of S-parameters which we use to a derive a predicted maximum possible WPTE along. Using the same framework, the dynamics of BS-RDAA is described in Section 4 where we show that the steady-state behavior of the system achieves the predicted maximum WPTE under a certain condition. Section 5 discusses the experiment done to prove the theoretical analysis in Section 4. Section 6 discusses the results and analysis of the data. Section 7 concludes this paper and outlines the future work that can be undertaken to improve upon the proposed BS-RDAA concept.

## 2. Overview of Power Beam Control of WPT Systems

The different methods proposed for the aforementioned purpose can be roughly categorized into three classes: 1) the conventional antenna array beam steering used in communication systems [7, 9, 13], 2) use of pilot signals to guide the generator antennas [14–16], or 3) feedback control using data sent back by the receiver to the generator [8, 9, 17]. Their respective block diagrams are shown in Figure 1.

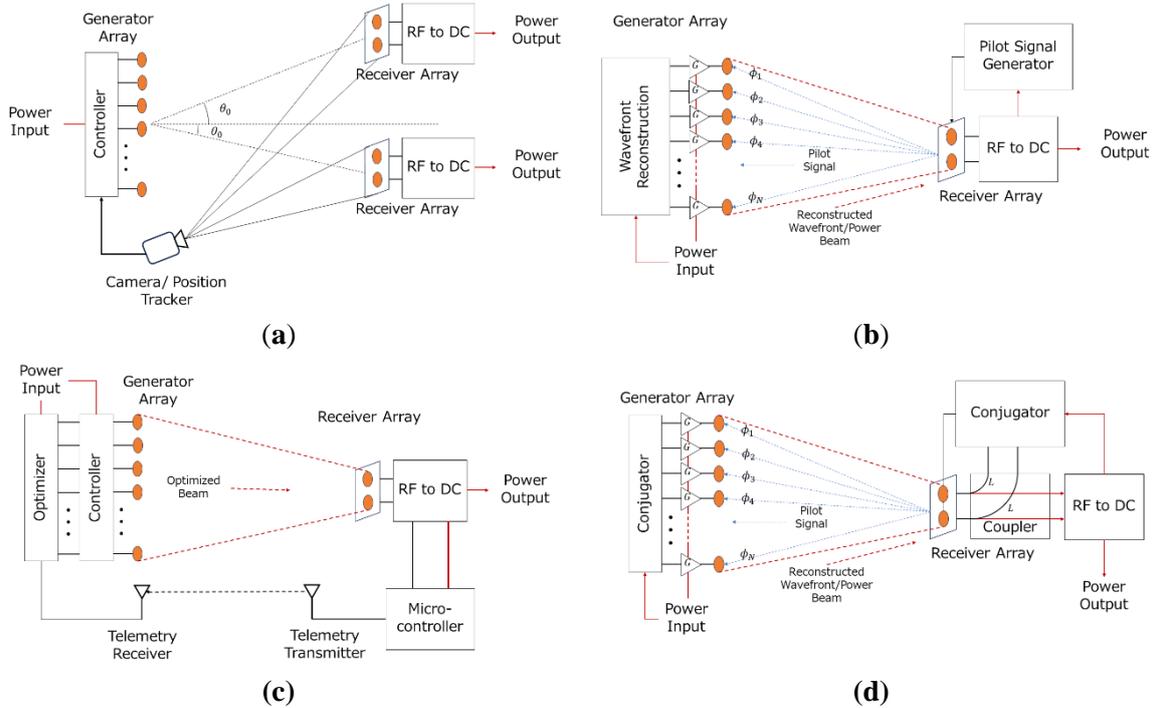

**Figure 1.** Block diagram of the different approaches for WPT beam control: (a) position tracking with conventional beam steering, (b) beam forming using pilot signals, (c) feedback control using telemetry data, and (d) the both-sides retrodirective antenna system.

2.1.1. Position tracking with conventional beam steering

The first type requires position tracking of the receiver through image recognition, radar, or satellite positioning. Once determined, the power generator will steer its beam towards the target using conventional beam forming techniques. Figure 1a roughly visualizes this system where image recognition using a camera is used to determine the position of the receivers. Sensors used in this system can also come in the form of GPS reporting, radar, or through a guidance signal [7].This process has a fast settling time but since the amplitude inputs are fixed and there is no knowledge of the received power, it can only achieve sub-optimal efficiency. It is also dependent on the accuracy of the sensor used. Precision can be increased by simultaneously using different sensors [18].

2.1.2. Wavefront reconstruction using a pilot signal

The second type improves upon the first by letting the receiver system send a pilot signal to the generator. This gives the task of wavefront reconstruction to the generator side which has been developed for MIMO communication systems. In a sense, this system uses the generator elements themselves as sensors to determine the position of the receivers [9]. The pilot signal acts as a guide for the generator to find the receiver. Figure 1b shows the block diagram of this method. An optimal method to implement this is through the use of retrodirective antennas [10, 11, 16]. These are purely analog systems that use phase conjugation that accurately reconstructs the wavefront of the pilot signal regardless of the position of the elements relative to the array.

WPTE of this method varies with the pilot signal characteristics. Any random pilot signal setting can lead to different values of efficiency and an optimum amplitude and phase setting on each receiver element can achieve the maximum [15]. Although for a dynamic channel, estimation must be done to find the optimum setting. This requires feedback and sensing which is in the scope of the third type of power beam control.

2.1.3. Feedback control using telemetry data

This family of approaches ensures that the power transferred is always maximized using feedback, commonly using the power measured at the receiver sent back as telemetry data. Figure 1c represents this method. The telemetry data enables the generator to infer the characteristics of the channel and solve an inverse problem that can find the optimum generator setting [17]. Other works use the data in an optimization problem loop which tunes the phase and/or amplitude parameters of all the antenna elements [8, 9].

Due to the feedback of data, the optimum setting can always be achieved by this type of power beam control even in a dynamic environment., due to the need for optimization and processing in the generator side, the complexity of the system and the convergence time increases as more elements are added into the system. This type of power beam control method becomes impractical as the WPT system grows in size.

2.1.4. Both-Sides Retrodirective Antenna Array (BS-RDAA) System

The BS-RDAA system [12] shown in Figure 1d combines the ideas behind the second and third type of power beam control. By installing retrodirective capability on the generator array and receiver array, a feedback loop is created. Continuous operation at marginal stability results in WPT at maximum efficiency with a faster settling time. Since it uses retrodirective arrays, it can scale up to larger systems consuming less power compared to digital processors and amplitude and phase controllers.

A caveat of using this technique is that marginal stability should be maintained which requires a gain control capability on the generator side or an attenuation control capability on the receiver side. However, compared to the optimization-based control problem from the third type, this approach is much simpler. The symmetry of receiver and generator implementation also adds flexibility to the design. For example, a WPT from a large generator to a small receiver entails that the gain control must be installed in the generator. contrast, the limited capability of the generator launched into space in SBSP entails that the ground receiver must have the gain control devices.

2.1.5. Comparison, contrast, and summary

On all the approaches, there is generally a trade-off observed between the WPTE and the convergence time needed to achieve their steady state. The first type has a fast convergence time but since there is no knowledge of the received power, it can only achieve sub-optimal efficiency. Use of pilot signals from the receiver as a form of guidance in the second type improves upon this but its WPTE is highly dependent on the pilot signal design. Its performance can be sub-optimal in the case when the pilot signal does not hit the generator which can likely happen due to the time-varying nature of the channel. The use of feedback in the third type ensures best WPTE, but the scanning process may take many iterations to converge which makes this approach slower than the other two. Furthermore, the computational complexity needed to perform optimization can become impractical.

A more detailed comparison of the different works on power beam control can be found in Table 1 in terms of the added hardware needed and a qualitative description of their performance. Generally, more capability is added in the generator side, and this may not be applicable especially to SBSP where the generator must be launched into space. The size, weight, and power requirements for each antenna element must be considered. In this regard, the BS-RDAA provides a more balanced solution in which it can achieve maximum WPTE while maintaining a faster convergence time and reduced computational complexity. Thus, we investigate this technique as a potential solution towards the problem of efficient long distance power beaming.

**Table 1.** Comparison of different works on wireless power beam control in terms of the needed receiver capabilities, generator capabilities, and their steady-state performance.

| Method | Type | Receiver Capabilities | Generator Capabilities | Performance |
|---|---|---|---|---|
| Visual tracking with camera [7] | Conventional beam steering | (none) | Visual tracking, precise phase control | Sub-optimal efficiency, Fast convergence |
| Directional radiation using position information [9] | Conventional beam steering | Position reporting | Position tracking, phase control | Sub-optimal efficiency, Fast convergence |
| Focused pilot signal transmission [15] | Pilot signal beam forming | Pilot signal transmission, Phase control | Retrodirective antenna array | Sub-optimal efficiency, Fast convergence |
| *In situ* antenna calibration [16] | Pilot signal beam forming | Pilot signal transmission, Phase control | Retrodirective phase control, digital processing | Sub-optimal efficiency, Slow convergence |
| Iterative power optimization [8] | Feedback control | Telemetry transmission | Processor capable of optimization, phase control | Sub-optimal efficiency, Slow convergence |
| Channel estimation from receiver power [17] | Feedback control | Telemetry transmission | Signal processor for channel estimation, phase control, amplitude control | Maximum efficiency, Slow convergence |
| Directional radiation by iterative superposition [9] | Feedback control | Telemetry transmission | Simple signal processing, phase control, amplitude control | Maximum efficiency, Slow convergence |

| Both-sides retrodirective antenna array [12] | Feedback control, Pilot signal beam forming | Retrodirective antennas, automatic gain control [a] | Retrodirective antennas, automatic gain control [a] | Maximum efficiency, Fast convergence |
|---|---|---|---|---|

[a] Only implemented on one side and never on both simultaneously.

## 3. Radiative Near-Field WPT Efficiency

An emerging approach to calculate WPTE draws inspiration from circuit theory by creating a circuit representation of the WPT channel. Since the system is MIMO with thousands to billions of ports, it is impractical to use resistors, inductors, and capacitors for this representation. A higher abstraction is used in the form of circuit network parameters [19] such as the Z-, Y-, and the S-parameters. For the theoretical analysis, we will only focus on the S-parameters which are widely used in high frequency circuits [20].

*3.1. Mathematical notation*

A scalar quantity is written in italic form, a column vector quantity is written in lower case bold font, and a matrix is written in upper case bold font. For example, $a$ and $A$ are scalars, **a** is a column vector, and **A** is a matrix. The notation $\{a_i\}$ indicates a column vector for a positive integer set defined by $i$ and $\{A_{ij}\}$ is a matrix with a column index $i$ and a row index $j$, both positive integers. Since these are complex quantities, we then define $|a|$ and $\angle a$ as the magnitude and phase of $a$, respectively.

The transpose and Hermitian transpose operations of **a** are defined by $\mathbf{a}^T$ and $\mathbf{a}^H$, respectively, and these operations transform a column vector into a row vector and vice versa. The per-element conjugation operation is defined by $\bar{\mathbf{a}}$ where each element of **a** is conjugated. The three operations outlined above can also be applied to matrices. The conjugation operation can also be done on scalars.

Since the dynamics of the BS-RDAA concept is discussed, we introduce the discrete-time index $(k)$. The value of some quantities $a$, **a**, or **A** at some time $(k)$ is denoted by $a^{(k)}$, $\mathbf{a}^{(k)}$, and $\mathbf{A}^{(k)}$, respectively. We differentiate this with the exponentiation notation $a^k$ where $a$ is a base and $k$ is an exponent.

We denote $\mathbb{R}$ and $\mathbb{C}$ as the set of all real numbers and complex numbers, respectively. Lastly, we use **I** to represent an identity matrix that matches the size of the matrix that is added to it.

*3.2. S-parameter model of efficiency*

S-parameters are used to characterize how a voltage wave is scattered in a microwave circuit with an arbitrary number of ports. It is a more accurate representation of all electric circuits, and it is widely used to characterize high frequency circuits. S-parameters have been used to characterize waveguides – devices that uses EM fields to propagate information [20]. The wireless channel can also be imagined as a waveguide without boundaries; thus, the S-parameter definition can be extended to antennas communicating with each other via the wireless channel. Figure 2a shows a simple two-antenna channel which can be related by a $2\times2$ S-parameter matrix. Both $v_1$ and $v_2$ have a forward (or going into the channel) and backward (or going out of the channel) wave components expressed as $v_1 = v_{1f} + v_{1b}$ and $v_2 = v_{2f} + v_{2b}$, respectively. The quantities are related by equation (1) where subscript $f$ means forward wave while $b$ means backward wave.

$$\begin{pmatrix} v_{1b} \\ v_{2b} \end{pmatrix} = \begin{pmatrix} S_{11} & S_{12} \\ S_{21} & S_{22} \end{pmatrix} \begin{pmatrix} v_{1f} \\ v_{2f} \end{pmatrix} \tag{1}$$

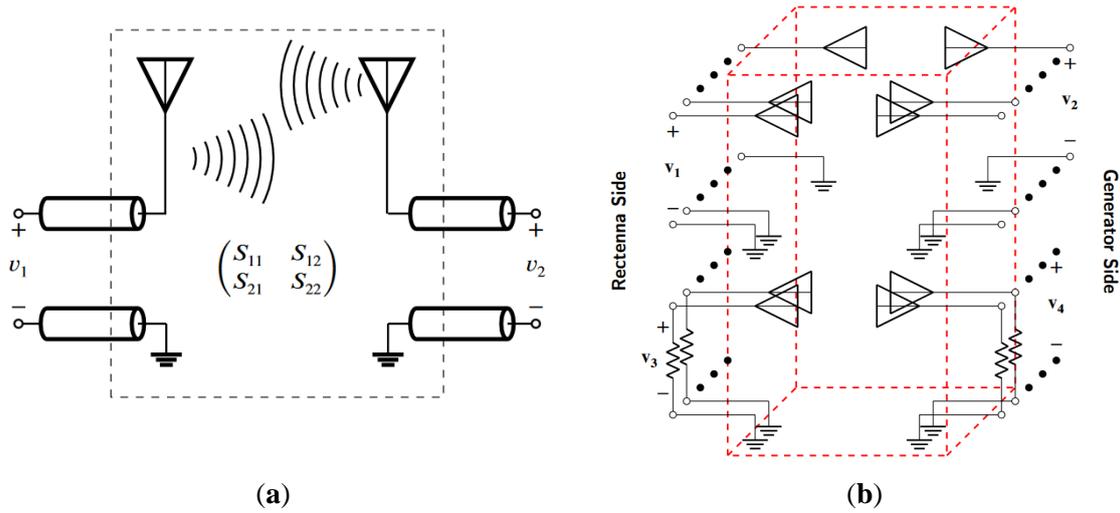

(a) (b)

**Figure 2.** (a) wireless channel between two antennas modelled by S-parameters, (b) WPT system model which is a multiple port version of the former.

Adopting the formalism from [19], the concept can be extended to $N$ power generating antennas and $M$ power harvesting antennas. A complete wireless channel view can be modelled by assuming an infinite number of ports where only the power generators and receivers are active as visualized in Figure 2b. Here, $\mathbf{v_i} = \mathbf{v_{if}} + \mathbf{v_{ib}}$ for $i \in \{1,2,3,4\}$, $\mathbf{v_1} \in \mathbb{C}^{M\times1}$, $\mathbf{v_2} \in \mathbb{C}^{N\times1}$, and $\mathbf{v_3}$ and $\mathbf{v_4}$ are infinitely large column vectors terminated by a matched load representing power absorbed by free space. Equation (2) can be used to model the scattering behavior of the system where $\mathbf{v_{3f}} = \mathbf{v_{4f}} = \mathbf{0}$.

$$\begin{pmatrix} \mathbf{v_{1b}} \\ \mathbf{v_{2b}} \\ \mathbf{v_{3b}} \\ \mathbf{v_{4b}} \end{pmatrix} = \begin{pmatrix} \mathbf{S_{11}} & \mathbf{S_{12}} & \mathbf{S_{13}} & \mathbf{S_{14}} \\ \mathbf{S_{21}} & \mathbf{S_{22}} & \mathbf{S_{23}} & \mathbf{S_{24}} \\ \mathbf{S_{31}} & \mathbf{S_{32}} & \mathbf{S_{33}} & \mathbf{S_{34}} \\ \mathbf{S_{41}} & \mathbf{S_{42}} & \mathbf{S_{43}} & \mathbf{S_{44}} \end{pmatrix} \begin{pmatrix} \mathbf{v_{1f}} \\ \mathbf{v_{2f}} \\ \mathbf{v_{3f}} \\ \mathbf{v_{4f}} \end{pmatrix} \tag{2}$$

Let $\mathbf{S}$ be the large S-parameter matrix from equation (2). Since the channel is lossless and reciprocal, then $\mathbf{S} = \mathbf{S}^T$ and $\mathbf{S}^H\mathbf{S} = \mathbf{SS}^H = \mathbf{I}$. Using this model, the location of each element, radiation properties of the antenna, mutual coupling, reflection losses, and channel conditions are all considered. From equation (2), using $\mathbf{v_{2f}}$ as the input, the WPTE can be expressed as in equation (3).

$$\eta = \frac{\mathbf{v_{2f}}^H(\mathbf{\bar{S}_{21}}\mathbf{S_{21}^T})\mathbf{v_{2f}}}{\mathbf{v_{2f}}^H\mathbf{v_{2f}}} \tag{3}$$

Equation (3) is the Rayleigh coefficient and it has been found in previous works when maximizing the WPTE of antenna arrays [4, 8] by treating $\mathbf{v_{2f}}$ as the input. Its maxima is the maximum eigenvalue of $\mathbf{\bar{S}_{21}S_{21}^T}$. Let $\xi$ be the eigenvalues of the matrix. Since $\mathbf{\bar{S}_{21}S_{21}^T}$ is Hermitian, then $\xi \in \mathbb{R}$ for all eigenvalues. Moreover, due to the unitarity of $\mathbf{S}$, and $\mathbf{S_{21}}$ being a submatrix of it, then the condition $|\xi| \leq 1$ holds true for all its possible values. Consequently, the maximum WPTE is governed by the largest eigenvalue, $\xi_{max}$.

We can represent an arbitrary vector input to the array using a weighted sum of the set of normalized eigenvectors, $\mathbf{a}_i$, of $\mathbf{\bar{S}_{21}S_{21}^T}$ as expressed in equation (4) where $v_{2i} \in \mathbb{C}$ for all $i \in \{1, 2, \ldots, M\}$. This assumes that $\mathbf{\bar{S}_{21}S_{21}^T}$ is a full rank matrix.

$$\mathbf{v_{2f}} = \sum_{i=1}^{M} v_{2i} \mathbf{a}_i \tag{4}$$

Now, the efficiency of this arbitrary input can be expressed in terms of the weights and eigenvalues as shown in equation (5). This also shows that each eigenvalue, $\xi_i$, is the WPTE of the beam mode, $\mathbf{a}_i$, and the efficiency is a weighted average of the individual efficiencies of each beam mode.

$$\eta = \frac{\sum_{i=1}^{M} \xi_i |v_{2i}|^2}{\sum_{i=1}^{M} |v_{2i}|^2} \tag{5}$$

A scalar multiple of the eigenvector associated with $\xi_{max}$, denoted by $\mathbf{a}_{max}$, is the optimum setting for the input. In the case that multiple eigenvectors are associated with $\xi_{max}$, then the optimum input can be a linear combination of the eigenvectors. The power can be controlled by the magnitude of the weights. Then, to achieve high efficiency WPT, a system must be designed such that the maximum WPTE condition, $\mathbf{v_{2f}} = c\mathbf{a}_{max}$, is maintained as the channel changes over time.

*3.3. Eigenvectors as components to determine efficiency*

Each eigenvector, $\mathbf{a}_i$, corresponds to a radiation pattern mode or beam mode of the generator antenna array that transfers power to the rectenna array, and is orthogonal to the other modes. The orthogonality indicates that each beam mode does not interfere with each other. This is analogous to the transmission modes of an antenna array with a significant coupling coefficient between elements [21], but instead of decomposing $\mathbf{S_{11}}$, the transmission S-parameter $\mathbf{S_{21}}$ is decomposed. An illustration of this notion is shown in Figure 3 where a four-element generator is simulated to transfer power to a four-element receiver. The power flow from the generator side is displayed on all cases. Each element is driven by a source corresponding to the magnitude and phase of each eigenvector from the resulting S-parameters of the simulation.

In the least eigenvalue case, Figure 3b, the magnitude of the Poynting vector hitting the receiver array is at a minimum as evident by the heatmap showing mostly dark areas. The next is Figure 3c where the power flow is increased due to the emergence of white areas.

Figure 3d shows more power going towards the receiver with more white areas illuminating the antenna elements. Lastly, in Figure 3e, the most efficient case where most of the receiver array is illuminated by the beam from the receiver as observed by a larger white area covering its heatmap.

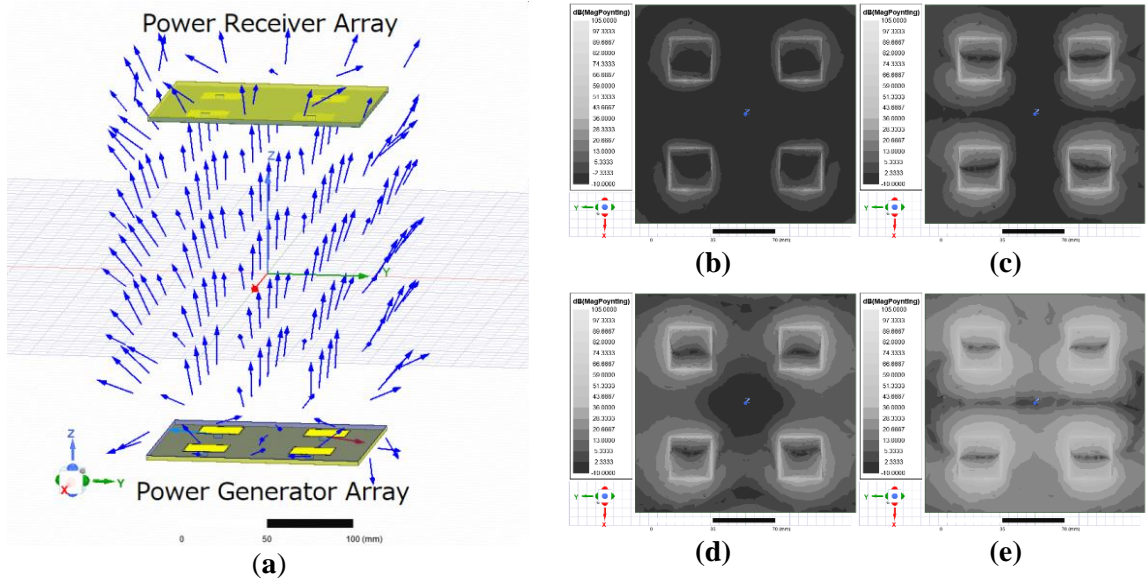

**Figure 3.** A simulation of a four-element power generator array sending power to a four-element receiver array: (a) Poynting vector field between the two arrays in the largest eigenvalue transmission mode, and (b)-(e) heat map of the magnitude of the Poynting vector on the receiver array plane arranged from smallest eigenvalue transmission mode to largest eigenvalue transmission mode.

## 4. Dynamics of the Both-Sides Retrodirective Antenna Array

The block diagram of the BS-RDAA system is shown in Figure 4, assuming negligible mismatch and mutual coupling between antennas on their respective arrays. From this, the discretized characteristic equation can be derived as equation (6), where one sampling period is assumed to be the round-trip time of the signal from the rectenna side to the generator side.

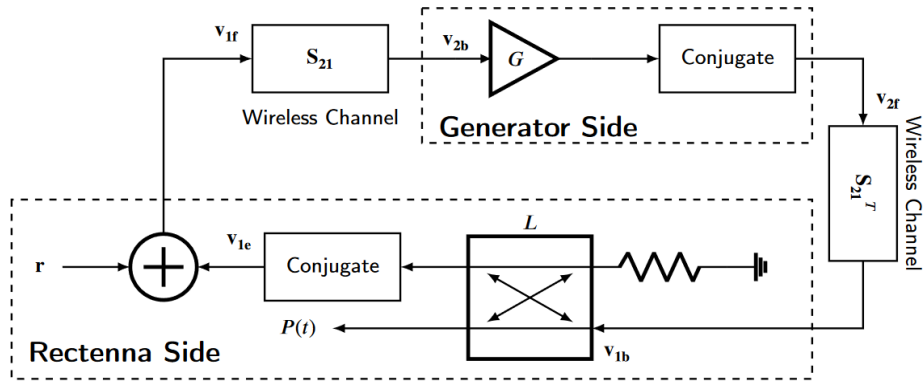

**Figure 4.** Block diagram of a both-sides retrodirective system with some input vector **r**.

$$\mathbf{v}_{1f}^{(k)} = \bar{L}G\mathbf{S}_{21}^H\mathbf{S}_{21}\mathbf{v}_{1f}^{(k-1)} + \mathbf{r}^{(k)}, \quad \mathbf{v}_{2f}^{(k)} = \bar{L}G\bar{\mathbf{S}}_{21}\mathbf{S}_{21}^T\mathbf{v}_{2f}^{(k-1)} + \bar{G}\bar{\mathbf{S}}_{21}\bar{\mathbf{r}}^{(k)} \quad (6)$$

Note that $\mathbf{S}_{21}^H\mathbf{S}_{21}$ and $\bar{\mathbf{S}}_{21}\mathbf{S}_{21}^T$ have the same set of eigenvalues. Without loss of generality, we assume $M < N$, and this also defines the maximum number of eigenvalues of the system. The zero-input response for an arbitrary initial condition is given by equation (7). Here, $\mathbf{b}_i$ is an eigenvector of $\mathbf{S}_{21}^H \mathbf{S}_{21}$ and $v_{1i}$ are weights for $\mathbf{v_{1f}}$, while $v_{2i}$ and $\mathbf{a}_i$ holds the same definition as that of equation (4). The set of eigenvectors, $\mathbf{b}_i$, is parallel to the physical aspect of $\mathbf{a}_i$, but it represents the beam modes from the rectenna to the generator.

$$\mathbf{v_{1f}}^{(k)} = (\bar{L}G)^k \sum_{i=1}^{M} \xi_i^k v_{1i}^{(0)} \mathbf{b}_i, \quad \mathbf{v_{2f}}^{(k)} = (L\bar{G})^k \sum_{i=1}^{M} \xi_i^k v_{2i}^{(0)} \mathbf{a}_i \qquad (7)$$

Marginal stability is achieved when the steady state value is finite and non-zero as $k$ approaches infinity. This is satisfied when the value $|LG|$ is the reciprocal of $\xi_{max}$ as expressed in equation (8). $|LG| > \xi_{max}^{-1}$ will render the system unstable while $|LG| < \xi_{max}^{-1}$ makes it stable. The resulting steady state expressions in marginal stability are given by equation (9).

$$|LG| = \frac{1}{\xi_{max}} \qquad (8)$$

$$\lim_{k \to +\infty} \mathbf{v_{1f}}^{(k)} = (\bar{L}G\xi_{max})^k v_{1,max}^{(0)} \mathbf{b}_{max}, \quad \lim_{k \to +\infty} \mathbf{v_{2f}}^{(k)} = (L\bar{G}\xi_{max})^k v_{2,max}^{(0)} \mathbf{a}_{max} \qquad (9)$$

Since steady state value of $\mathbf{v_{2f}}$ is a scalar multiple of $\mathbf{a}_{max}$, computing the efficiency according to equation (3) results to $\eta = \xi_{max}$. Therefore, achieving marginal stability on BS-RDAA achieves maximum WPTE. To maintain marginal stability, $L$ or $G$ must change as $\mathbf{S_{21}}$ changes over time. This will require automatic gain/attenuation control which is outside the scope of this study.

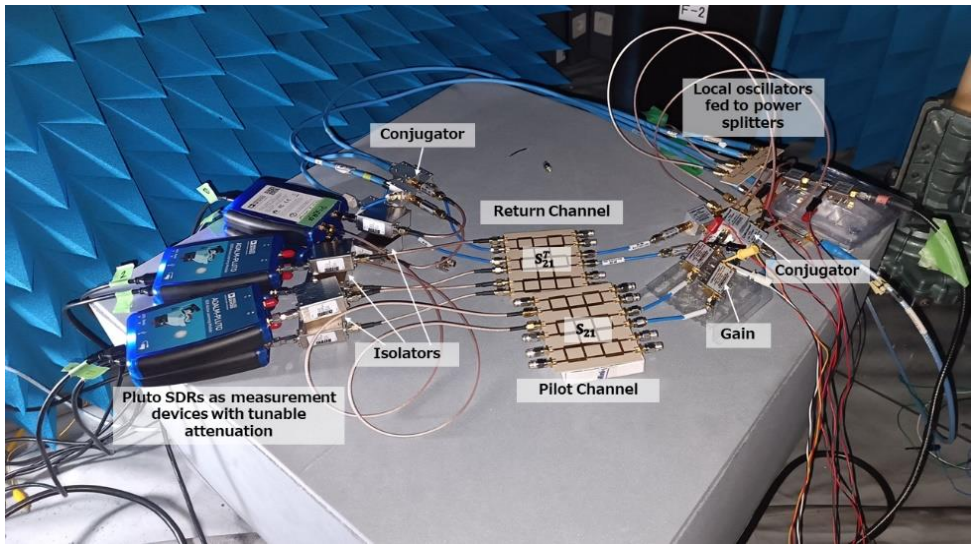

**Figure 5.** Hardware experiment setup.

## 5. Experimentation

To validate the preceding theoretical analysis, we implement the experiment hardware setup shown in Figure 5. A twelve-port channel model is used for this experiment to make S-parameter measurements easily repeatable and for further data processing. Knowing the S-parameters of the channel is required to calculate the maximum efficiency predicted by equation (3). Two identical implementations of this channel model enable the representation of $\mathbf{S}_{21}$ and $\mathbf{S}_{21}^T$, respectively. Modularized microwave components such as power splitters, amplifiers, isolators, and mixers are used to assemble the whole experiment setup to resemble the model in Figure 4.

The block diagram of the proposed experiment is shown in Figure 6 for a setup that uses ports 1 and 2 for $\mathbf{v}_1$ and ports 7 and 8 for $\mathbf{v}_2$. The return channel sends the power from the generator ports and is scattered to ports 1 to 6 where power is measured by the software-defined radios (SDR) connected to a host computer. The active receiver ports are connected to a conjugating circuit first before their power is measured and subsequently, the conjugated signal is retransmitted by their corresponding SDR to the pilot channel. The internal transmit gain of the SDR is varied to simulate the unstable, marginally stable, and stable conditions predicted by equation (8). After the pilot channel, the signals experience a fixed gain $G$ and a conjugation before being looped back to the return channel.

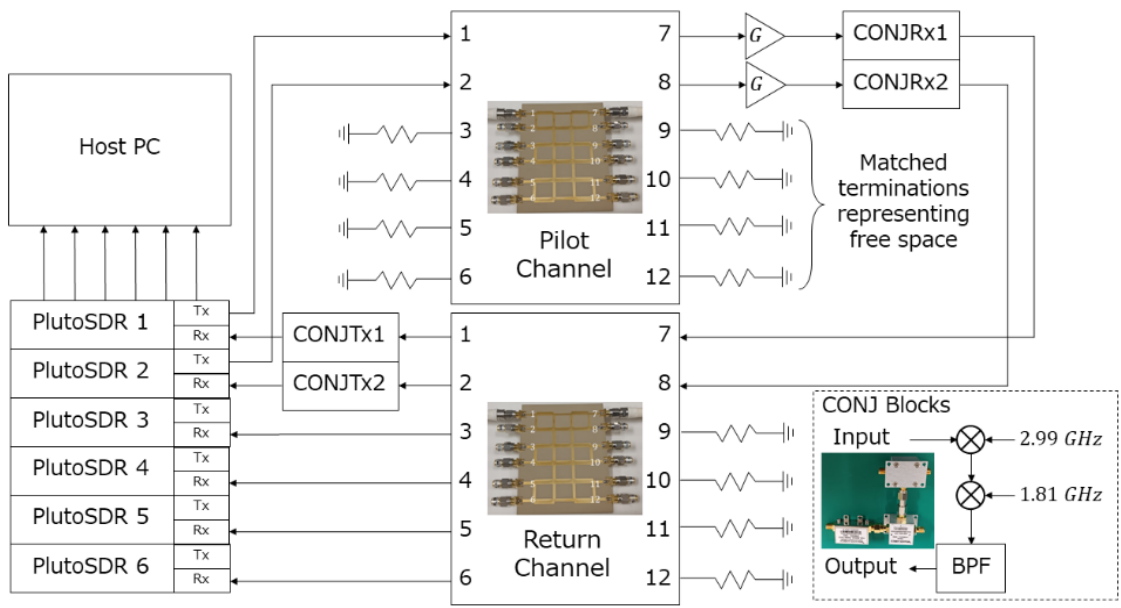

**Figure 6**. Block diagram of the proposed experiment setup to validate the theoretical analysis of the dynamics of BS-RDAA. This particular diagram shows the case when ports 1 and 2 are activated at the receiver side while ports 7 and 8 are activated at the generator side.

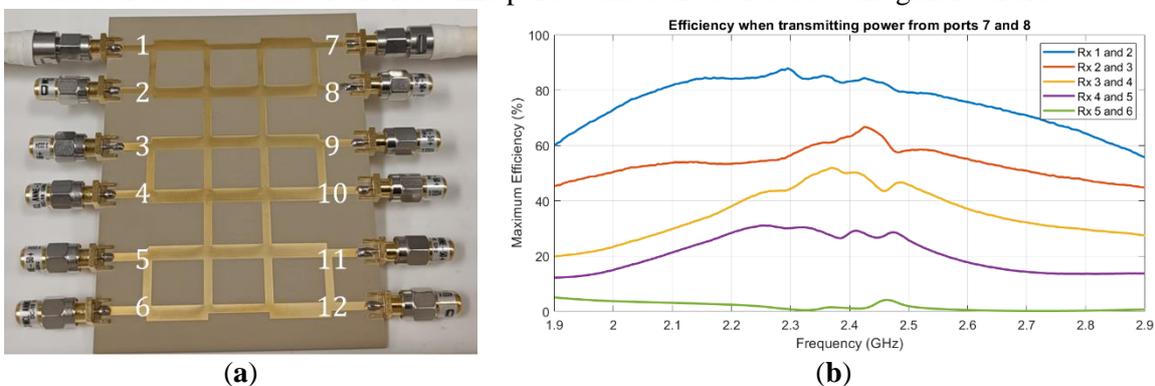

**Figure 7.** Fabricated 12-port channel model: (a) fabricated hardware with the port numbers labelled, (b) maximum efficiency when transmitting power from port 7 and port 8 to some combinations of ports 1 to 6.

## 5.1. Channel model

Figure 7a is the channel model fabricated on a Rogers TC600 laminate and it was designed for a center frequency of 2.4 GHz. It consists of multiple directional couplers that makes ports 1 to 6 isolated from each other and transfers most of their power to ports 7 to 12. The same behavior is also observed in the reverse direction. With the correct amplitude and phase fed to ports 1 to 6, the power can be focused on any port from 7 to 12, similar to the dynamic beam focusing concept. By connecting a matched load on a port, we can model the power lost to the wireless channel as that port will absorb any power it receives.

The channel model circuit board is designed in reference to the S-parameter view of WPT as shown in Figure 2b. Ports 1 to 6 represent $\mathbf{v_1}$ and $\mathbf{v_3}$ on the side of the receiver array and ports 7 to 12 represent $\mathbf{v_2}$ and $\mathbf{v_4}$ on the side of the generator array. By choosing the active ports on both sides, $\mathbf{v_1}$ and $\mathbf{v_2}$ are decided while all inactive ports are terminated by a matched load so that any signal arriving at those ports are purely absorbed. The absorbing ports now represent $\mathbf{v_3}$ and $\mathbf{v_4}$.

Using the information from the S-parameter measurements, we can determine the best ports to be used for the highest efficiency, similar to antennas directly facing each other. We can also simulate a channel that has poor wireless power transfer efficiency by choosing the ports that do not transmit much energy towards each other. For example, using ports 7 and 8 to transmit power to ports 1 and 2 is a high efficiency wireless channel while using the same ports to transmit to ports 5 and 6 simulates a low efficiency wireless channel. The maximum eigenvalue of $\mathbf{S}_{21}^H \mathbf{S}_{21}$ can be used to predict the value of the maximum efficiency and this is shown in Figure 7b where ports 7 and 8 are transmitting power to different combinations of ports 1 to 6. As the receiver moves from ports 1 and 2 to ports 5 and 6, a gradual decrease in efficiency is observed and this can represent the SBSP satellite rotating on its own axis relative to the ground receiver. Transmitting to ports 1 and 2 can mean that the satellite is directly facing the ground receiver while transmitting to ports 5 and 6 can represent that satellite facing 90° away from the direction of the ground receiver.

## 5.2. Conjugating circuit

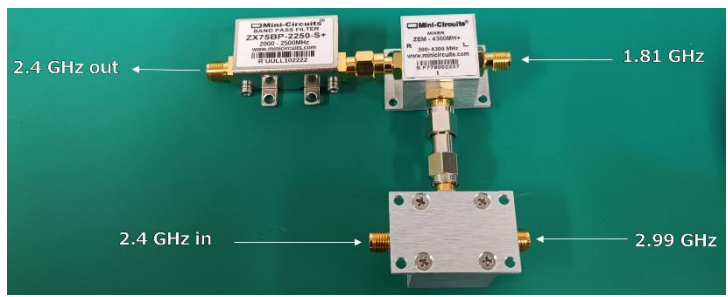

**Figure 8**. Hardware implementation of the superheterodyne conjugating circuit.

The core of retrodirective capability is the use of conjugating circuits. For this experiment, we use the superheterodyne implementation which is also the implementation modelled in the time domain simulation. The conjugating circuit, whose implementation is shown in Figure 8, has a higher frequency of 2.99 GHz and a lower frequency of 1.81 GHz ensuring the same center frequency of operation. We originally used 3 GHz and 1.8 GHz, but spurious intermodulation products occupied the 2.4 GHz band interfering with measurements. Note that 2.4 GHz is a fourth-order intermodulation product of 3 GHz and 1.8 GHz. This implementation of the conjugating circuit uses passive mixers, so attenuation is expected. It is measured to be 16.5 dB in this implementation, and this must be factored into the efficiency measurement calculation.

*5.3. SDR measurement and retransmission algorithm*

Six (6) Pluto SDRs are used to measure the power received on each port in the receiver side. The active SDRs not only measure the signal power but also retransmit them on their corresponding transmit port connected to the forward channel without making any changes to the amplitude and phase. This simplifies the implementation while keeping the closed feedback loop. In addition, the gain can be internally controlled, and the spectrum of each radio can be tracked on a user interface programmed in GNU Radio.

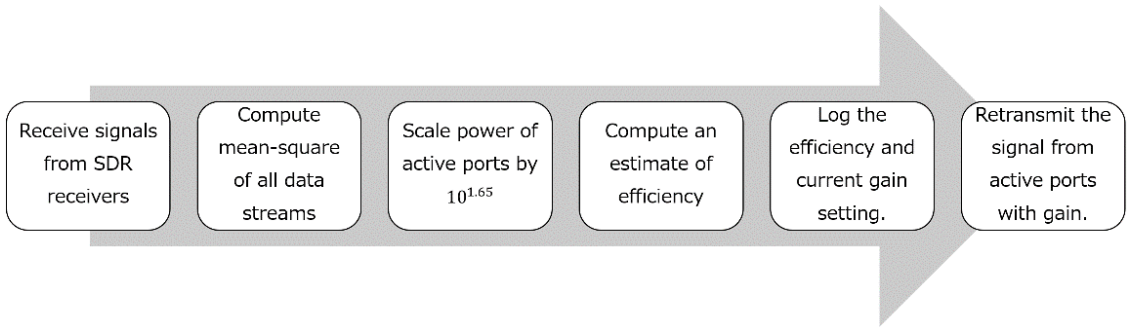

**Figure 9**. Flow of the program running on the Pluto SDRs that measure the power at the receiver ports, make an estimate of the efficiency, log the data, and retransmit the signals in the active ports.

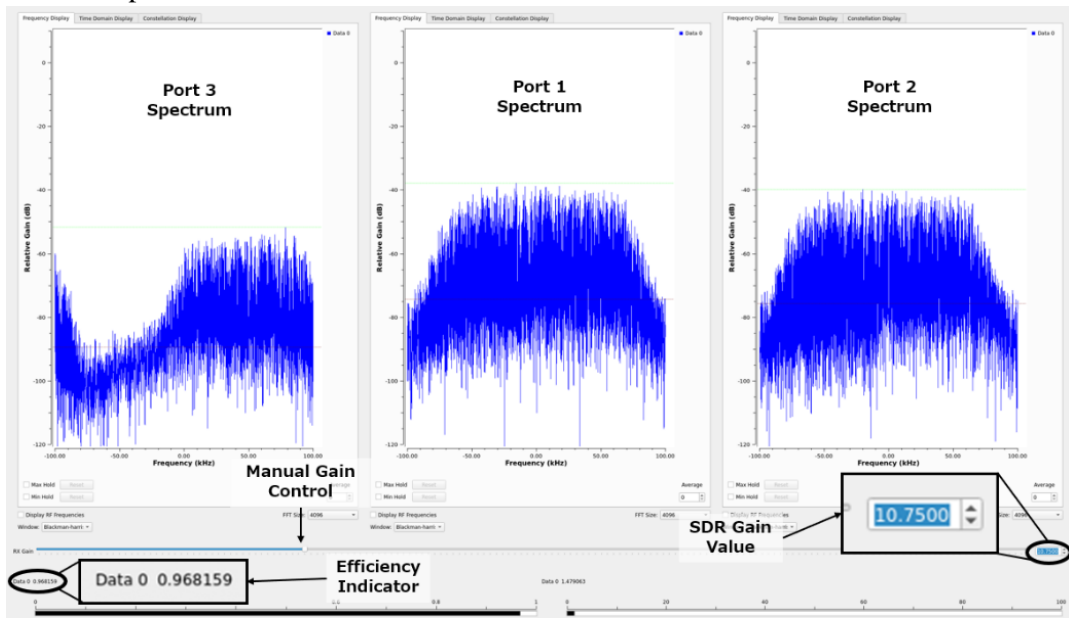

**Figure 10**. Measurement indicators and loop gain control interface using GNU Radio. The spectra illustrate marginal stability condition.

Figure 9 shows the flowchart of the SDR program used for the experiment. The program simultaneously takes the data streams from all the SDRs and measures the power received on each port through the mean-square value. Due to the loss from the conjugating circuits connected before the SDRs at the active ports, the power measured from them must be scaled accordingly. After scaling, the program makes an estimate of the efficiency and logs this data to a file for further processing. Finally, the signal received at the active ports are retransmitted with some gain imposed by the program's current setting.

Figure 10 shows the user interface at a case when the system is in marginal stability. In this interface, the gain can be manually set by the user. This mechanism is used to vary the loop gain of the system and observe the unstable, marginally stable, and stable conditions predicted from the theoretical analysis. An estimate of the efficiency can also be seen in this interface.

*5.4. Measurement methodology*

The 12×12 S-parameter matrix of the channel board in Figure 7a is measured using a vector network analyzer (VNA). To see the behavior of the system stability, the gain value is swept from high to low. Theory predicts that the system moves from the unstable condition to the stable condition from this sweep method. A sharp change in measured efficiency is observed as the gain sweeps from higher to lower gain. This is visually confirmed when the spectrum displayed in Figure 10 becomes mostly noise indicating a transition from instability to stability. Marginal stability can be found in this transition region and is noted for all test cases that were examined. In each gain point, the efficiency is logged by the SDR host computer with a sampling rate of 40 samples per second. The mean and standard deviation of the collection of efficiencies at each gain point is calculated.

Alongside the efficiency value, we note the gain point at marginal stability for further statistical analysis using linear regression. Thirty (30) combinations of transmit ports (Tx ports 7-12) and receive ports (Rx ports 1-6) are tested. The influence of the Pluto SDR gain tuning is also factored in after the statistical analysis to obtain a more accurate linear regression result. This is largely influenced by the gain tuning of the Pluto SDR device and its effect is characterized by using a loop back test and performing a linear regression to verify the slope of the tuning. Ideally as the gain setting in decibels is changed, we also see the same change in the measured power by the SDR in decibel units also, i.e., they are linearly related with a slope of $m = 1$.

Efficiency is calculated using the power measured at the Rx ports 1 to 6 using equation (10) where $P_i$ is the power measured at the $i^{th}$ port. Measurement at the power side is avoided so that the SDRs can be set to receive the same power level avoiding signal saturation as they can only receive up to 100 mW. In effect, there needs to be some compensation factor to measure the real efficiency, and this is estimated from the S-parameter measurements of the board using the VNA.

$$\eta_{meas} = \frac{10^{1.65}\alpha \sum_{i=active} P_i}{P_1 + P_2 + P_3 + P_4 + P_5 + P_6}, \quad \alpha = \frac{\mathbf{v}_{2f}^H(\mathbf{S}_{12}^H\mathbf{S}_{12} + \mathbf{S}_{13}^H\mathbf{S}_{13})\mathbf{v}_{2f}}{\mathbf{v}_{2f}^H \mathbf{v}_{2f}} \tag{10}$$

The $10^{1.65}$ factor compensates for the loss experienced by the signal when passing through the conjugating circuit at the rectenna side. A factor $\alpha$ is multiplied to factor in the power lost due to the inactive Tx ports, radiation losses, and ohmic losses by the circuit board implementation. It is calculated based on the measured S-parameters as defined by equation (2) where $\mathbf{v}_3$ in Figure 2b represents the inactive ports measured at the rectenna side of the circuit board in Figure 7a. The vector $\mathbf{v}_{2f}$ that is used to calculate $\alpha$ comes from the resulting eigenvector of the characteristic equation in (6).

### 5.5. Statistical analysis

To verify the theoretical prediction, it is not enough to show that the theoretical efficiency matches the measured efficiency from the experiment. Further validation can be found by looking at the marginal stability condition in equation (8). Particularly, we investigate the decibel expression of this as expressed in equation (11) where $L_{dB} = -20\log_{10}(L)$ and $G_{dB} = 20\log_{10}(G)$.

$$20\log_{10}(\eta_{max}) = -G_{dB} + L_{dB} \tag{11}$$

In the experiment setup, due to internal settings inside the SDR, the total loop loss, $L_{dB}$, is difficult to measure. Only the gain setting $G_{dB}$ and $\eta_{max}$ can be measured by the experiment setup. By measuring multiple cases (i.e., different combinations of ports 1 to 6 and ports 7 to 12) on the same setup, different values of $\eta_{max}$ can be measured, all with different $G_{dB}$ setting. Performing linear regression on the measured data points should result into the linear relationship between $20\log_{10}(\eta_{max})$ and $G_{dB}$ in equation (11). It should result in a slope of $m = -1$ with a constant $L_{dB}$.

Another test can be done by computing an estimate of $L_{dB}$. Given a data point with $\eta_{max}$ measured at a set $G_{dB}$, an estimate of the loss can be made. The average of this quantity can be used as an estimate for $L_{dB}$ and the resulting curve can be used to check the fit of the data. The coefficient of determination shown in (12) is used to evaluate the fit of the two curves to the data set and verify the prediction made by the analysis in the previous section.

$$R^2 = 1 - \frac{\sum_{i=1}^{K}(y_i - mG_{dB,i} - L_{dB})^2}{\sum_{i=1}^{K}(y_i - \langle y \rangle)^2} \tag{12}$$

In the equation above, letting $y = 20\log_{10}(\eta_{max})$, then $y_i$ is the measured efficiency value, $G_{dB,i}$ is the corresponding gain for $y_i$, $L_{dB}$ is the estimated loss value, $m$ is the slope which is computed in the linear regression curve and set to $m = -1$ in the loss estimation curve, $\langle y \rangle$ is the mean of the data set, and $K = 30$ is the total number of data points measured. If $R^2 > 0.8$, then the experiment setup validates the proposed model for the behavior of the BS-RDAA concept. Ideally, $R^2 = 1$ but the 0.2 deficit can be attributed to compounding errors in the methodology and measurement noise.

## 6. Results and Discussion

### 6.1. Determining the point of marginal stability

Thirty (30) different combinations of receiver ports 1 to 6 and generator ports 7 to 12 were tested. The numerical results are tabulated in Table 2 in the Appendix A. Figure \ref{fig:test-case-results} show the graphical result of the efficiency measurements under gain variation for cases 1 to 3 showing how the point of marginal stability was determined. The gain starts at 20 dB and is manually swept step-by-step to 5 dB. We can observe a sharp change in measured efficiency as we decrease the gain. It is the region where the marginal stability point occurs. Also notice that at higher maximum efficiency, the transition points also occur at lower values of the gain. This confirms the condition $|LG| = 1/\xi_{max}$ required to achieve marginal stability as higher eigenvalues require lower gain or attenuation.

In the stable condition, on each test case, we can observe a measurement floor where the efficiency measured is dictated by the noise power at each SDR. In the unstable condition, as the gain is increased, there is a noticeable decreasing trend as gain increases. This is due to the effect of other eigenvalues becoming more present which reduces the total efficiency as described by equation (5).

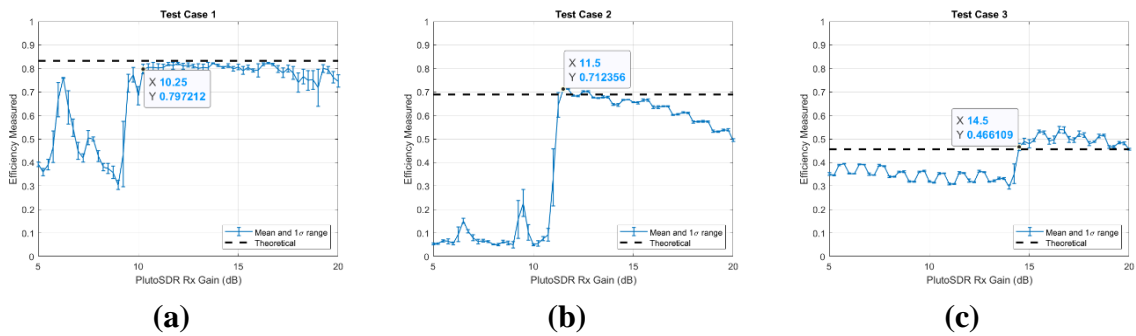

**(a)** **(b)** **(c)**

**Figure 11**. Experiment results from three test cases with a one standard deviation error bar for each gain point (solid blue lines) with the point of marginal stability highlighted and compared to their respective maximum efficiency predicted by theory (dashed black lines).

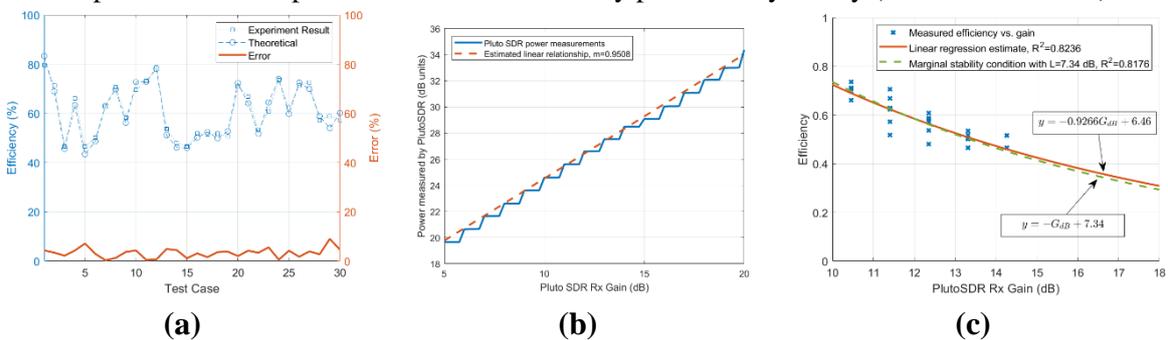

**(a)** **(b)** **(c)**

**Figure 12**. Result of further analysis on gathered data: (a) testing the Pluto SDR gain linearity, (b) error of measured efficiency compared to theory, and (c) linear regression performed on the data to verify the marginal stability condition.

### 6.2. Validation of data gathered

Figure 12a shows the comparison of the theoretical expectation for the efficiency and the measured efficiency from the experiment. The theoretical efficiency is calculated using equation (3) from the S-parameter measurements of the channel board, assuming that $\mathbf{v_{2f}} = \mathbf{a}_{max}$. Measurement error between the theoretical and measured values is computed by $100\% \times |\eta_{theo} - \eta_{meas}|/\eta_{theo}$ where $\eta_{theo}$ is the theoretical value while $\eta_{meas}$ is the measured value. The resulting error values for all test cases are within 10% which is acceptable. Thus, the experiment demonstrates the ability of the BS-RDAA concept to achieve maximum WPTE predicted by theory.

Further validation of the preceding theoretical analysis is done by confirming the correctness of the marginal stability condition. First, the linearity of the Pluto SDR gain tuning is tested. The result is a stepping behavior is observed in Figure 12b where the gain tuning is only observed to change at 1 dB steps instead of 0.25 dB. Thus, the floor value of the gain at marginal stability is noted as the point of marginal stability. Further, an estimated slope of 0.9508 is seen for every 1 dB change in the setting. Therefore, the noted gain values for all test cases are adjusted accordingly.

The resulting linear regression curve is shown in Figure 12c. We determine that the constant loss is $L_{dB} = 6.46\ dB$ and the slope is $m = -0.9266$ which results into the red line from the plot. The predicted slope from theory is $m = -1$ so the linear regression result has a 7.34% error which is acceptable. The resulting coefficient of determination for this is $R^2 = 0.8236$ which is greater than 0.8 indicating a strong agreement between theory and experiment.

The green curve from Figure 12c represents the line defined by equation (11) where $L_{dB}$ is calculated as the mean of the last column from Table 2. The slope is assumed to be $m = -1$ in this case. The coefficient of determination of this curve to the data set is $R^2 = 0.8176$ which also serves as further proof that the experiment setup validates the theory.

### *6.3. Possible sources of measurement error*

A source of measurement error include the error in measuring the scaling factor in equation (10) where the 16.5 dB loss from the conjugating circuit is compensated in the efficiency measurement calculation. The theoretical results and the factor $\alpha$ are highly dependent on the S-parameter measurements using the two-port VNA where systematic errors can build up as more ports are measured. Another source can be due to an error on the model itself and further refinement can be done. Better measurement methods and/or use of better instruments can reduce the errors between experiment and theory that arose from the experiment setup detailed in this paper. Nonetheless, the former affirms the latter and that the BS-RDAA system can achieve the maximum theoretical WPTE.

### **7. Conclusions and Recommendations**

Control of long distance WPT systems becomes a herculean task as the distance grows larger requiring larger antennas. Different techniques have been developed to achieve a high efficiency system. In this paper, we investigated the concept of the both-sides retrodirective antenna array system which shows a good balance between a fast convergence time and power transfer efficiency. We presented a discrete-time state-space model to characterize its dynamics and reveal that the system can achieve maximum theoretical efficiency as long as marginal stability is maintained. Further verification of the proposed model is performed through a hardware experiment centered on a 12-port channel model board that simulates a static wireless channel. Different channel conditions can be made by different combinations of active and inactive ports from which thirty test cases were measured. The measured efficiency is within 10% of the expected efficiency. The marginal stability condition predicted by equation (8) is also examined where the $R^2$ values for the linear regression and using the data to estimate $L$ are both greater than 0.8 which indicates that the theory is in good agreement with the data gathered from the experiment.

Further refinement of either the model or the experiment setup is needed to have a better measurement result when compared to the theoretical prediction. For future work, we are looking into refining the experiment setup to use better measuring equipment and move to a wireless channel instead of a channel model board. A control system to maintain marginal stability of the proposed system is also a topic for further research. Preferably, a simple approach of automatic gain control using techniques from Classical Control Theory is attractive for a large-scale system.

**Acknowledgments:** The hardware experiment presented in this article is partially funded by Hitachi, Ltd. through their partnership with Prof. Shinichi Nakasuka and the Intelligent Space Systems Laboratory (ISSL) in The University of Tokyo. Through the experiment, we were able to move from simulation and confirm our theoretical analysis in a real world implementation.

**Conflicts of Interest:** Naoki Shinohara is an editor-in-chief and Bo Yang is an associate editor for Space Solar Power & Wireless Transmission and were not involved in the editorial review or the decision to publish this article. All authors declare that there are no competing interests. The funders had no role in the design of the study, in the collection, analyses, or interpretation of data, in the writing of the manuscript, or in the decision to publish the results.

**Appendix A: Experiment Data**

Table 2. Theoretical efficiency, measured efficiency, error, estimated gain setting, and estimated loss for all thirty (30) test cases.

| Case | Rx Ports | Tx Ports | Efficiency (%) | | Error (%) | Gain (dB) | | Est. Loss (dB) |
|---|---|---|---|---|---|---|---|---|
| | | | Theoretical | Measured | | Setting | Corr. | |
| 1 | | 7, 8 | 83.29 | 79.72 | 4.29 | 10 | 9.51 | 7.54 |
| 2 | 1, 2 | 8, 9 | 68.91 | 71.24 | 3.38 | 11 | 10.46 | 7.51 |
| 3 | | 9, 10 | 45.63 | 46.61 | 2.15 | 14 | 13.31 | 6.68 |
| 4 | | 7, 8 | 63.31 | 66.01 | 4.26 | 11 | 10.46 | 6.85 |
| 5 | 2, 3 | 9, 10 | 43.40 | 46.49 | 7.12 | 15 | 14.26 | 7.61 |
| 6 | | 10, 11 | 48.62 | 50.05 | 2.94 | 14 | 13.31 | 7.30 |
| 7 | 1, 2 | 7, 9 | 63.02 | 62.82 | 0.31 | 12 | 11.41 | 7.37 |
| 8 | | 8, 10 | 69.79 | 70.66 | 1.24 | 12 | 11.41 | 8.39 |
| 9 | | 7, 11 | 56.18 | 58.25 | 3.68 | 13 | 12.36 | 7.67 |
| 10 | 1, 3 | 8, 10 | 72.68 | 69.56 | 4.30 | 11 | 10.46 | 7.31 |
| 11 | 1, 4 | 8, 9 | 72.87 | 73.20 | 0.46 | 10 | 9.51 | 6.80 |
| 12 | | 8, 10 | 78.13 | 78.64 | 0.65 | 10 | 9.51 | 7.42 |
| 13 | | 10, 12 | 51.13 | 53.64 | 4.91 | 13 | 12.36 | 6.95 |
| 14 | 3, 4 | 9, 10 | 46.10 | 48.12 | 4.39 | 13 | 12.36 | 6.01 |
| 15 | | 8, 9 | 45.96 | 46.47 | 1.12 | 14 | 13.31 | 6.65 |
| 16 | | 7, 8 | 50.14 | 51.70 | 3.11 | 14 | 13.31 | 7.58 |
| 17 | 3, 4 | 7, 9 | 52.38 | 51.56 | 1.57 | 15 | 14.26 | 8.51 |
| 18 | | 7, 10 | 49.91 | 51.71 | 3.61 | 12 | 11.41 | 5.68 |
| 19 | | 10, 12 | 52.71 | 50.71 | 3.80 | 14 | 13.31 | 7.41 |
| 20 | 2, 3 | 7, 9 | 72.29 | 70.83 | 2.02 | 11 | 10.46 | 7.46 |
| 21 | | 7, 10 | 64.09 | 66.78 | 4.19 | 12 | 11.41 | 7.90 |
| 22 | | 8, 10 | 51.76 | 53.49 | 3.35 | 14 | 13.31 | 7.88 |
| 23 | 2, 3 | 7, 11 | 64.45 | 60.86 | 5.57 | 13 | 12.36 | 8.05 |
| 24 | 2, 4 | 7, 10 | 73.95 | 73.55 | 0.54 | 11 | 10.46 | 7.79 |
| 25 | | 8, 11 | 59.80 | 62.31 | 4.20 | 12 | 11.41 | 7.30 |
| 26 | 1, 6 | 10, 11 | 72.44 | 71.19 | 1.73 | 11 | 10.46 | 7.51 |
| 27 | | 8, 9 | 69.93 | 72.65 | 3.90 | 10 | 9.51 | 6.73 |
| 28 | | 7, 12 | 58.83 | 57.24 | 2.70 | 13 | 12.36 | 7.51 |
| 29 | 2, 5 | 7, 8 | 53.98 | 58.83 | 8.99 | 13 | 12.36 | 7.75 |
| 30 | | 11, 12 | 60.11 | 57.33 | 4.62 | 12 | 11.41 | 6.58 |